\documentclass[reprint,aps,prl,superscriptaddress,showpacs,floatfix,preprintnumbers,longbibliography]{revtex4-1}
\usepackage{graphicx}
\usepackage{color}
 
\usepackage{booktabs}
\usepackage{multirow}

\usepackage[utf8x]{inputenc}

\definecolor{linkcol}{rgb}{0.2,0.2,0.2}

\usepackage[pdfencoding=auto,psdextra]{hyperref}
\usepackage{bookmark}
\def\BAIO{Ba$_5$AlIr$_2$O$_{11}$}
\def\JF{$J_{\rm eff}$}

\begin{document}

%\preprint{DRAFT 10}

%\title{Competing spin-orbit coupling and orbital hybridization effects in mixed valence  iridate Ba\texorpdfstring{\textsubscript{5}}{\texttwoinferior}AlIr\texorpdfstring{\textsubscript{2}}{\texttwoinferior}O\texorpdfstring{\textsubscript{11}}{\texttwoinferior}} 

\title{Charge ordering in Ir dimers in the ground state of Ba\texorpdfstring{\textsubscript{5}}{\texttwoinferior}AlIr\texorpdfstring{\textsubscript{2}}{\texttwoinferior}O\texorpdfstring{\textsubscript{11}}{\texttwoinferior}}

%\title{Spin-orbit coupled ground state of mixed valence iridate Ba\texorpdfstring{\textsubscript{5}}{\texttwoinferior}AlIr\texorpdfstring{\textsubscript{2}}{\texttwoinferior}O\texorpdfstring{\textsubscript{11}}{\texttwoinferior}}
\author{Vamshi M. Katukuri}
\altaffiliation{Current addres: Max Planck Institute for Solid State Physics, Heisneberstrasse 1, 70569, Stuttgart Germany}
\email{V.Katukuri@fkf.mpg.de}
\affiliation{Institute of Physics, \'{E}cole Polytechnique F\'{e}d\'{e}rale de Lausanne (EPFL), CH-1015 Lausanne, Switzerland}

\author{Xingye Lu}
\email{luxy@bnu.edu.cn}
\affiliation{Swiss Light Source, Photon Science Division, Paul Scherrer Institut, CH-5232 Villigen PSI, Switzerland}
\affiliation{Center for Advanced Quantum Studies and Department of Physics, Beijing Normal University, Beijing 100875, China}

\author{D. E. McNally}
\affiliation{Swiss Light Source, Photon Science Division, Paul Scherrer Institut, CH-5232 Villigen PSI, Switzerland}

\author{Marcus Dantz}
\affiliation{Swiss Light Source, Photon Science Division, Paul Scherrer Institut, CH-5232 Villigen PSI, Switzerland}

\author{Vladimir N. Strocov}
\affiliation{Swiss Light Source, Photon Science Division, Paul Scherrer Institut, CH-5232 Villigen PSI, Switzerland}

\author{M. Moretti Sala}
\affiliation{European Synchrotron Radiation Facility, BP 220, F-38043 Grenoble Cedex, France}

\author{M. H. Upton}
\affiliation{Advanced Photon Source, Argonne National Laboratory, Argonne, Illinois 60439, USA}

\author{J. Terzic}

\author{G. Cao}
\affiliation{Department of Physics and Astronomy, University of Kentucky, Lexington, Kentucky 40506, USA}
\affiliation{Department of Physics, University of Colorado at Boulder, Boulder, CO 80309}

\author{Oleg V. Yazyev}
\affiliation{Institute of Physics, \'{E}cole Polytechnique F\'{e}d\'{e}rale de Lausanne (EPFL), CH-1015 Lausanne, Switzerland}

\author{Thorsten Schmitt}
\email{thorsten.schmitt@psi.ch}
\affiliation{Swiss Light Source, Photon Science Division, Paul Scherrer Institut, CH-5232 Villigen PSI, Switzerland}

\begin{abstract} 
It has been well established  experimentally that the interplay of electronic correlations and spin-orbit interactions in Ir$^{4+}$ and Ir$^{5+}$ oxides results in insulating \JF=1/2 and \JF=0 ground states, respectively. 
However, in compounds where the structural dimerization of iridum ions is favourable, the direct Ir $d$--$d$ hybridisation can be significant and takes a key role. 
Here, we investigate the effects of direct Ir $d$--$d$ hybridisation in comparison with electronic correlations and spin-orbit coupling in \BAIO, a compound with Ir dimers. 
Using a combination of {\it ab initio} many-body wave function quantum chemistry calculations and resonant inelastic X-ray scattering (RIXS) experiments, we elucidate the electronic structure of \BAIO. 
We find excellent agreement between the calculated and the measured spin-orbit excitations.
Contrary to the expectations, the analysis of the many-body wave function shows that the two Ir (Ir$^{4+}$ and Ir$^{5+}$) ions  in the Ir$_2$O$_9$ dimer unit in this compound preserve their local \JF\ character close to 1/2 and 0, respectively.
The local point group symmetry at each of the Ir sites assumes an important role, significantly limiting the direct $d$--$d$ hybridisation. 
Our results emphasize that minute details in the local crystal field (CF) environment can lead to dramatic differences in electronic states in iridates and 5$d$ oxides in general.  
\end{abstract}

%\pacs{}
\date{\today}
\maketitle

 Dimerization or clustering of TM atoms is observed in many TM compounds, e.g. in vanadium oxides~\cite{Imada_RMP_1998,LiVo2_PRL_1997} and titanates~\cite{khomskii_spinels_prl_2005} where dimers of spin singlets akin to the Peierls state in one dimension~\cite{Peierls_1991_book} are stabilized when the $t_{2g}$ orbitals of the TM $d$-manifold are partially filled.  
 In these systems, the TM ions tend to have a strong direct (intra-dimer) $d$--$d$ overlap that result in molecular-like orbitals with appreciable bonding-antibonding splitting. 
Consequently, the local electronic structure depends on the intra-dimer hopping integral ($t_d$), intra-atomic Hund's coupling ($ J_{H} $) and inter-atomic ($ U $) Coulomb interactions and electron filling of the orbitals localized at TM clusters.
Alternatively, dimerization of TM ions can also be favourable from crystallographic considerations, particularly in compounds with heavy TM ions, e.g. 5$d$ ions, where the $d$ orbitals are more spread out. 
A number of dimerized or cluster 4$d$ and 5$d$ compounds~\cite{Terasaki_bairo3_crystals_2016,Nag_d4_dimer_PRL_2016,Dey_IrDimer_PRB_2017,Nag_d4_PRB_2018} with intriguing properties have been synthesized recently.
 Novel physical phenomena have been observed in these compounds, e.g. the inelastic X-ray scattering analogue of Young's double slit experiment has been realized in Ba$_3$CeIr$_2$O$_9$~\cite{revelli_resonant_2019}, where the molecular orbital formation within the Ir dimers is crucial.
In lacunar spinels Ga$M_4$$X_8$ ($M$=Nb, Mo, Ta and W and $X$=S, Se and Te), spin-orbit coupled molecular $J_{\rm eff}$ states~\cite{kim_spin-orbital_2014,jeong_direct_2017} and topological superconductivity~\cite{park_pressure-induced_2020} have been proposed  where molecular orbital formation within the tetrahedral cluster of $M$ ions is the key. 
  
  The interplay of inter-site electron hopping ($t$), $J_{H} $, $U$ and the strong atomic spin-orbit coupling (SOC) in $5d$ and in some $4d$ compounds
result in the \JF\ physics~\cite{BJKim_PRL_2008,BJKIM_Sci_2009,katukuri_PRB_2012,moretti_sala_resonant_2014,Witczak_ARCMP_2014}.
For instance, in compounds with Ir$ ^{4+} $ ($d^5$) configuration in an octahedral environment, e.g. in Sr$ _{2} $IrO$ _{4}$~\cite{BJKim_PRL_2008,BJKIM_Sci_2009}, the strong SOC leads to completely filled \JF\ = 3/2 and half-filled \JF\ = 1/2 levels.
Similarly,  in Ba$_2$YIrO$_6$ and NaIrO$_3$, the Ir$ ^{5+} $ ions realise a completely filled \JF = 3/2 and empty \JF = 1/2 sub-manifolds~\cite{book_abragam_bleaney,Khaliullin_d4_PRL_2013}, resulting in a non-magnetic $J_{\rm eff} = 0$ ground state~\cite{Kush_d4_PRB_2018}.
In dimerized systems, $t_{d}$ can be much larger and successively may play a dominant role compared to other local interactions, 
which could result in the breakdown or a significant modification of the \JF\ physics. 
Thus, it is crucial to identify the role of these multiple physical interactions in Ir dimer systems to gain a better understanding of the electronic and magnetic properties of these materials.

In this letter, we illustrate how subtle crystal structure details are extremely important to precisely understand the electronic structure of 5$d$ compounds where structural dimerization or clustering is prevalent.
Using state-of-the-art {\it ab initio} many-body electronic structure methods in combination with high-resolution resonant inelastic X-ray scattering (RIXS) experiments, we present a detailed analysis of the electronic structure of Ir$_2$O$_9$ dimers in \BAIO and unravel the nature of electronic ground and excited states of \BAIO.  
While we find an excellent agreement between the RIXS spectra and the calculated excitations, analysis of the many-body wave functions reveal a nearly complete charge separation -- Ir$^{4+}$ and Ir$^{5+}$-- within the dimers in the ground state, in contrary to an earlier report of formation of molecular orbitals in \BAIO~\cite{BIAO_wang_prl_2019}.
The strong SOC of the Ir$^{4+}$ and Ir$^{5+}$ ions results in \JF = 1/2 and \JF = 0 local configurations, respectively, and thus we conclude that a localized \JF\, picture is more appropriate in \BAIO.
 
 \BAIO\, contains dimers composed of crystallographically  inequivalent Ir cations encaged in face sharing O$ _{6} $ octahedra~\cite{Buschbaum_BAIO_synthesis_1989,Terzic_BAIO_synthesis_PRB_2015},
 see Fig.~\ref{fig_energy_levels}a and \ref{fig_energy_levels}b,  and Supplementary material (SM) Fig.~S1 ~\cite{supmat}.  
 At 210 K, a lattice distortion is believed to lower the symmetry of the crystal and enhance the charge disproportionation leading to charge ordering that correspond to Ir$ ^{4+} $ and Ir$ ^{5+} $ valence configurations~\cite{Terzic_BAIO_synthesis_PRB_2015}. 
However, analysis of RIXS spectrum of \BAIO\  using density functional theory and model Hamiltonian calculations~\cite{BIAO_wang_prl_2019} has proposed the formation of hybridized dimer orbitals, debunking the charge disproportionation phenomenon.
Nevertheless, given the complex low-symmetry crystal environment and the interplay of spin and orbital degrees of freedom in \BAIO, it is unclear if the dimer orbitals are actually realised in the ground state.
\begin{figure}[!t]	
	\centering
 	\includegraphics[trim=0cm 0.0cm 0cm 0cm, clip=true, width=1.02\linewidth]{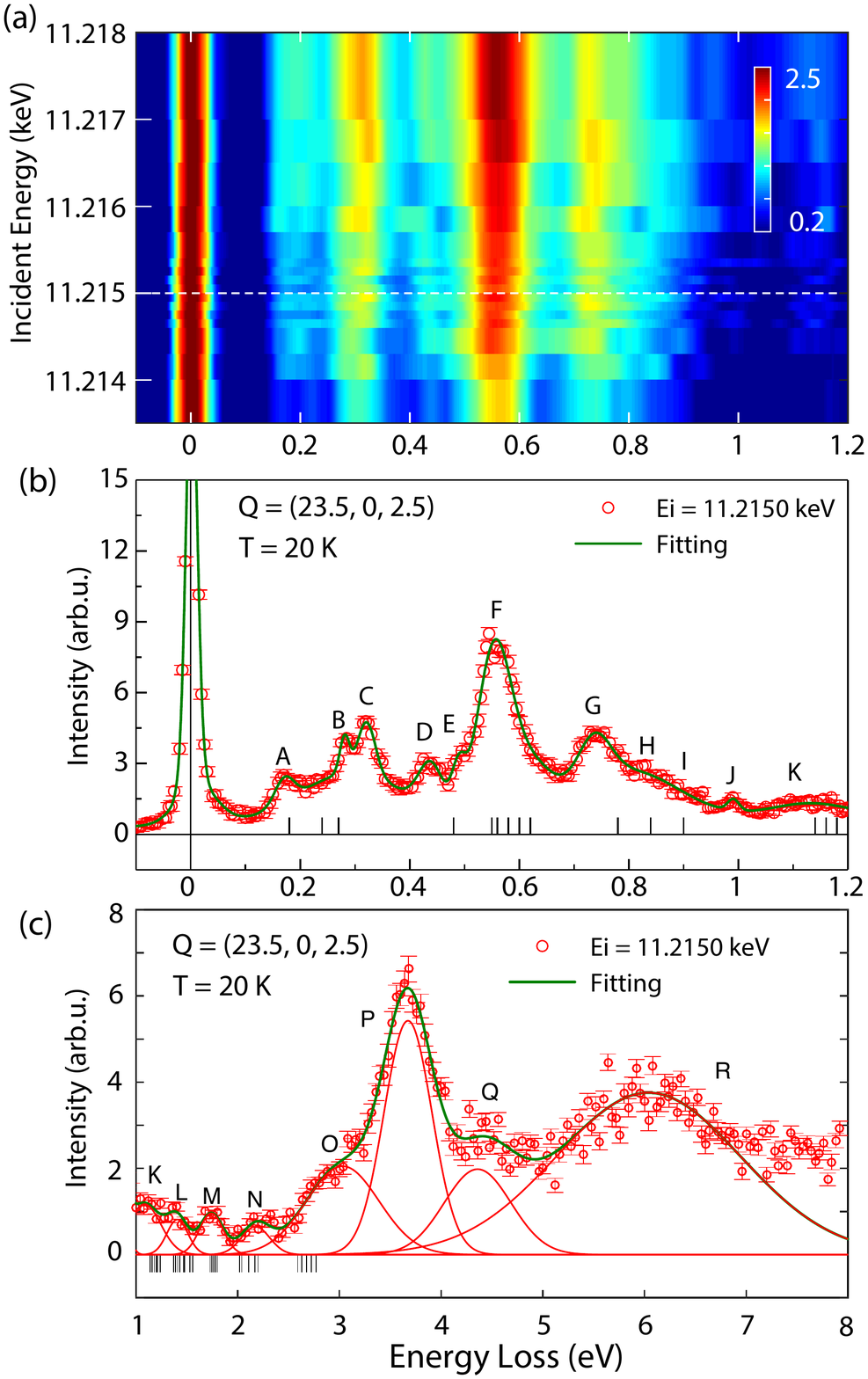}
	\caption{(a) Incident-energy dependence of elementary excitations below 1.2 eV for \BAIO\ measured around Ir $L_3$ edge with $ Q=(23.5, 0, 2.5) $ at 20 K. (b) RIXS spectra (below 1.2 eV) measured at  $E_i=11.215$ keV (marked as white dashed line in (a)). (c) High energy excitations (1-8 eV) measured with the same setup as that for (b). Green line is a Multi-Gaussian fitting of the raw data in red open circles (with error bars). 
		%The excited state energies computed from the QC calculations are shown with vertical bars in black. 
	%{\cred Vamshi, please add the vertical bars, or you can send me the values you want to add.}
	}
	\label{fig:dcirl}
\end{figure}

{\it Results}: The RIXS spectra shown in  Fig.~\ref{fig:dcirl} was measured on single crystals grown by flux method~\cite{Terzic_BAIO_synthesis_PRB_2015} at the ID20 beam line of the European Synchrotron Radiation Facility (ESRF) with $\sim$25 meV resolution~\cite{MorettiSala_ID20} and the 27-ID-B beamline with  $\sim$30 meV resolution at the Advanced Photon Source (APS), with $\pi$ polarization at a scattering angle close to $2\theta=90^\circ$.  
  The incident-energy dependence of RIXS spectra across the Ir-$L_3$ edge ($E_i=11.215$keV) at  the zone center Q=(23.5, 0, 2.5) is shown in Fig.~\ref{fig:dcirl}(a). 
   While the same $E_i$ as determined from previous measurements on iridates such as Sr$_2$IrO$_4$ and Ba$_2$YIrO$_6$~\cite{Jungho_PRL_2012,Kush_d4_PRB_2018} was chosen, we find that the maximum of the resonance is not at $E_i$ in \BAIO\ as the precise CF environment around Ir ions and the mixing of the valence states influences the resonance energy. 
	However, we see that the energies of the modes remain unchanged in a broad range around $E_i$.

The features marked by A-K in Fig.~\ref{fig:dcirl}(b) are incident-energy independent Raman modes as shown in Fig.~\ref{fig:dcirl}(a). These modes correspond to intrinsic electronic transitions between various occupied and unoccupied states, and therefore provide direct information about low energy electronic structure.
 To resolve all the Raman modes and determine the low energy electronic structure,  
we show in Figs.~\ref{fig:dcirl}(b) and ~\ref{fig:dcirl}(c) high statistic energy spectra collected at $E_i$ (white dashed line in Fig.~\ref{fig:dcirl}(a)). 
 In Fig.~\ref{fig:dcirl}(b), several sharp Raman modes below 1 eV and a broad peak at 1.2 eV, named $ A $ to $ K $ are  determined by fitting the spectra using multiple gaussians. The sum of the fitting curves is shown as a green solid curve. 
%In order to understand the energy levels above these, we present in 
In Fig.~\ref{fig:dcirl}(c), higher energy excitations up to 8 eV are shown. 
This spectrum is decomposed into several peaks and interestingly, these modes show very little momentum dependence (see SM Fig. S6 ~\cite{supmat}), indicating that all of them correspond to local spin-orbital ($ d-d $) excitations and reflecting the low energy electronic structure. 

\begin{table}[!b]	
	\caption{Relative energies (eV) of the excitation levels calculated at CASSCF+NEVPT2 level of theory. 
		The first column contains non-relativistic multiplet structure, the multiplet symbols on the left correspond to the octahedral ($ O_{h} $) symmetry. 
		The degeneracy of the states is split in \BAIO\ due to the lowered symmetry in the two octahedra due to the anisotropic crystal fields, see text. 
		Spin-orbit coupled multiplet structure is shown in the second column. Note that each state is  doubly degenerate (Kramers doublet). 
		%The degeneracy of each state is shown in the brackets, $D$: Doublet, $Q$: Quartet and $S$: Sextet. 
		The corresponding peaks in the RIXS data in Fig.~\ref{fig:dcirl}a are shown in column 3. 
		%The wave function for each of the states is also provided. 
	}
	\label{table:qc_wfc}
	%\begin{ruledtabular}
	\begin{tabular}{|ll||l|l|}
		\hline
		\hline
		
		\multicolumn{2}{|c||}{CASSCF+NEVPT2} & + SOC (x 2)& Ir L-edge RIXS \\
		
		\hline

		$ ^4 A_{1}$ -- &0.00                    		& 0.00 			     	  &0.00  \\[0.2cm]
		$ ^2 T_{1}$ -- &0.03, 0.08, 0.10   		 &0.18  					 & 0.18 (A)\\[0.02cm]
		$ ^2 A_{1}$ -- &0.14   					 	   &0.24  , 0.27        			& 0.28 (B) 0.33 (C) \\[0.2cm]
		%	&					        						 & 0.27                      &  0.33 (C)   \\	[0.02cm]
		$^4 T_{1}$ -- &	0.16, 0.17, 0.17    &                             &  0.44 (D)   \\	[0.12cm]
		$ ^2 E_{1}$ -- &0.18, 0.23                 &0.48, 0.62                     &  0.50 (E)   \\	[0.02cm]
		$ ^4 E_{1}$ -- &0.25, 0.28               &0.55, 0.56, 0.58  & 0.56 (F)\\[0.02cm]
		$ ^2 T_{2}$ -- &0.77, 0.80, 0.94     & 0.60, 0.78          &    0.75 (G)     \\[0.2cm] 
		
		$ ^4 T_{2}$ -- &0.84, 0.86, 0.95    & 0.84, 0.90                    &  0.82 (H) \\[0.02cm]
		
		$ ^2 T_{3}$ -- & 0.86, 0.86,0.90    & 1.14 -- 1.20 (4) & 0.98 (I), 1.00 (J) \\[0.2cm]
		
		$ ^2 A_{2}$ -- &	0.91			        &  1.21, 1.24          & 1.20 (K)    \\	[0.02cm]
		$ ^2 E_{2}$ -- & 1.00, 1.01              &  1.37 -- 1.43 (4) &    1.4 (L) \\	[0.02cm]
		$ ^2 A_{3}$ --&	1.03		               &   1.47  &  			    \\	[0.02cm]
		$ ^2 T_{4}$ -- &1.10,1.10,1.12        &   1.48, 1.53, 1.56           &  				   \\	[0.2cm]
		
		$ ^2 E_{3}$ --&	1.60, 1.63	           &    1.73 -- 1.80 (5)                 &  	1.77 (M)			  \\	[0.2cm]
		
		$ ^2 T_{5}$ --&1.68, 1.78, 1.81     &    2.02, 2.04                &  				   \\	[0.02cm]
		&					                  &  2.11, 2.17, 2.20         &  	2.17 (N)			   \\	[0.02cm]		                       
		&					                 &  2.59                               &  				   \\	[0.02cm]		                       
		&					                 &  2.63 -- 2.77 (4)        &  	 2.70	(O)		   \\	[0.02cm]
		%		                       &					          &  2.11, 2.17, 2.20         &  	{\cred 2.26} O-$K$			   \\	[0.02cm]		                       
		%		                       &					          &  2.59                       &  				   \\	[0.02cm]		                       
		%		                       &					          &  2.63 -- 2.77 (4)     &  	{\cred 2.71}	O-$K$		   \\	[0.02cm]		                       		
		\hline
		\hline
	\end{tabular}
	
	%	\end{ruledtabular}
\end{table}

\begin{figure}
	\centering
 	\includegraphics[trim=0cm 16.35cm 0cm 0cm, clip=true, width=0.95\linewidth]{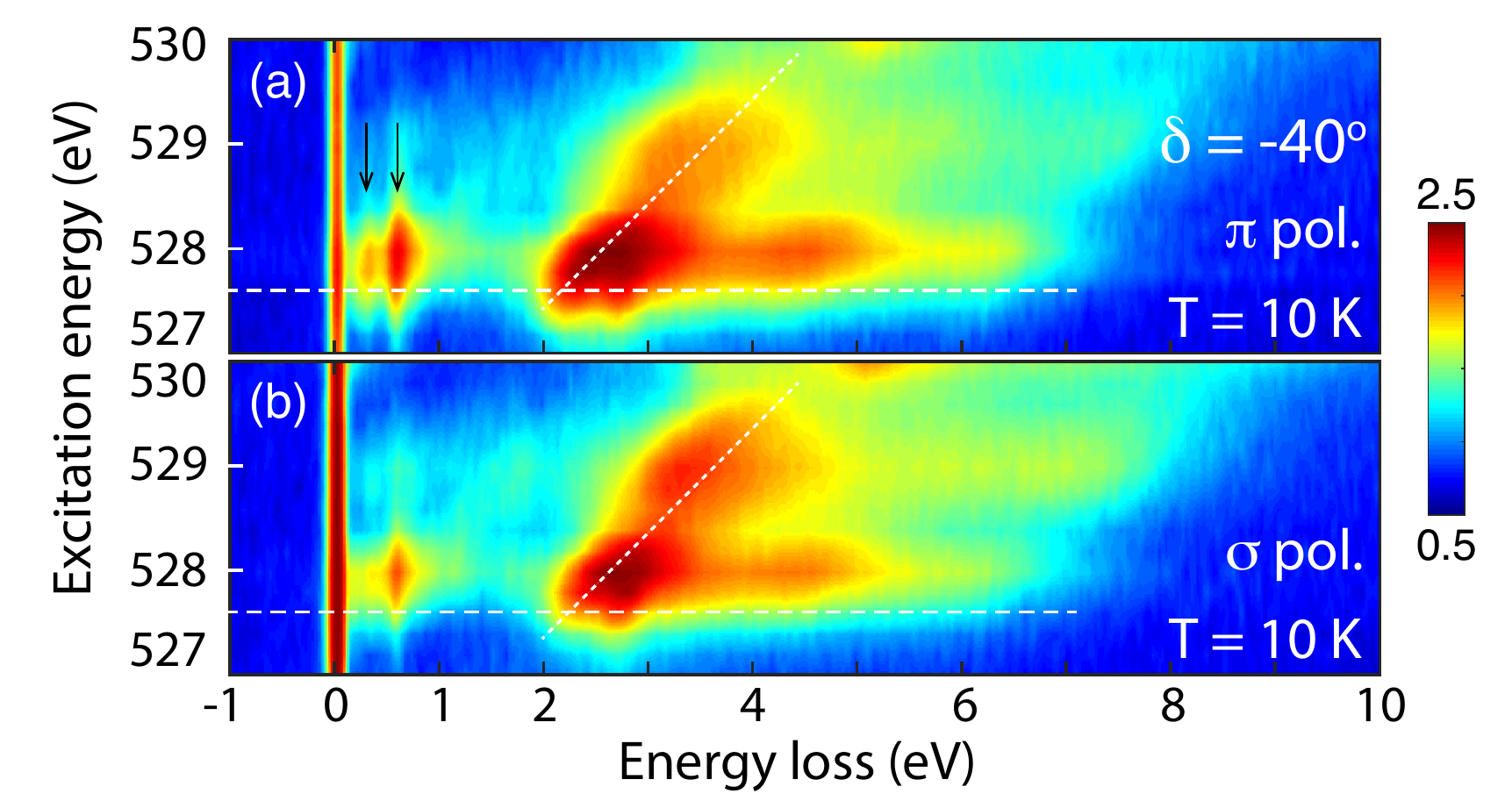}
	\caption{RIXS results of \BAIO\ as measured at O-$K$ edge. (a, b) Energy dependence of RIXS spectra for \BAIO\ taken around O-$K$ edge with $\delta=-40^\circ$  ($ Q=(0.93, 0, 0.65) $).
	%	 (c-f) High statistic energy spectra measured at pre-edge (527.6 eV). The insets show the scattering geometries. 
	 }
	\label{fig:OK}
\end{figure}

\begin{figure}[!t]
 	\includegraphics[trim=0.1cm 0.1cm 0cm 0cm, clip=true, width=0.98\linewidth]{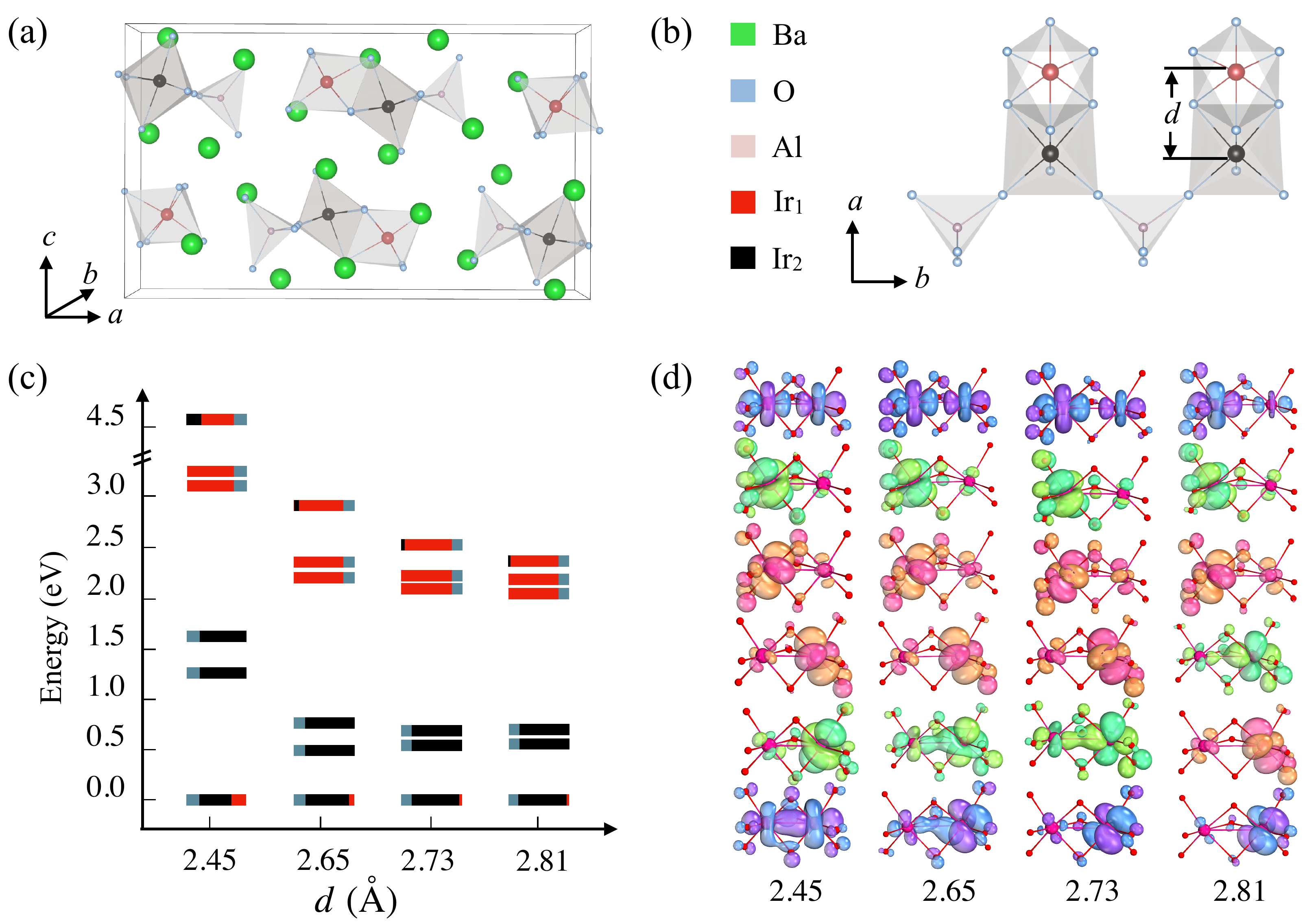}
 	\caption{Crystal structure of \BAIO\  : (a) unit cell, (b) Ir$ _{2} $O$ _{9} $ dimer units connected along $b$. $d$ is the intra-dimer Ir$_1$-Ir$_2$ distance. (c) Orbital (relative) energy-level diagram of the six $ t_{2g} $ orbitals, shown in (d), for different $d$, the lowest energy orbital is set to zero. The length of red, black and blue colours of each level are proportional to the percentage contributions from Ir$_{1}$, Ir$_{2}$ and O ions, respectively. %(c) Comparison of the spin-orbit coupled excitation energies for different $d$.  The average energy of each multiplet is plotted, labels $D$, $Q$ and $ S $ correspond to doublet, quartet and sextet multiplets respectively. See Supplementary table ~\cite{supmat} for exact energies.
 	}
 	\label{fig_energy_levels} 
 \end{figure}

We now turn to the RIXS results measured using O-$K$ edge  (Fig.~\ref{fig:OK} and Fig.~S2 in SM~\cite{supmat})  carried out at the ADRESS beamline of the Swiss Light Source at the Paul Scherrer Institut, with $\sim$70 meV energy resolution for both $\sigma$ and $\pi$ polarizations at a scattering angle of 2$\theta$=130 deg.~\cite{slspsi_beamline, saxes_2006}, see SM, Fig.~S1.
With the presence of strong hybridization between O $2p$ orbitals and Ir $5d$ orbitals, O-$K$ RIXS is sensitive to various elementary excitations of iridates~\cite{XingyeLu_OKedge_214_227_PRB_2018}. 
% Similar to the incident-energy dependent RIXS map measured around Ir $L_3$ edge, 
 Figs. 2(a) and 2(b) are RIXS maps collected at O-$K$ edge with $\pi$ and $\sigma$ polarizations at $25^\circ$ grazing incidence. 
 Besides the sharp spin-orbital excitations ($E\approx0.26, 0.57$ eV) below $1$ eV consistent with those measured with Ir-$L_3$ edge, two high energy excitations at $E\approx2.26$ and $2.71$ eV have also been observed. 
 Note the RIXS maps in Fig. 2 contains substantial fluorescence which is absent in the results collected at Ir $L_3$ edge (Fig.~\ref{fig:dcirl}), indicating complex energy levels/bands of oxygen ions.  
 In addition, significant polarization dependence of the excitations have also been observed, 
 %This can be attributed to overlaps between light polarization (electric field ${\bf E}$) and different O $2p$ orbitals hybridized with different Ir $5d$ orbitals (for details, see \cite{supmat}).
 which we attribute to the overlap between light polarization (electric field ${\bf E}$) and the different O $2p$ orbitals hybridized with different Ir $5d$ orbitals (for details, see~\cite{supmat}).

To decipher the nature of the rich excitation spectrum observed in RIXS spectra and to examine the formation of dimer orbitals in \BAIO, we performed many-body {\it ab initio} cluster-in-embedding quantum chemistry (QC) calculations, starting from the crystal structure reported in Ref.~~\cite{Buschbaum_BAIO_synthesis_1989,Terzic_BAIO_synthesis_PRB_2015}. 
These are based on the construction of the exact wave function for the atoms in the cluster using configuration interaction wave function theory --  complete active space self-consistent field  (CASSCF) and multireference perturbation methods~\cite{book_qc_2000}. 
The calculations were performed on a cluster containing one Ir$ _{2} $O$ _{9} $ dimer unit, two neighboring AlO$ _{4} $ tetrahedra and the surrounding 15 Ba$ ^{2+} $ ions. All the calculations were performed using ORCA quantum chemistry program~\cite{orca_4_neese}, see  SM~\cite{supmat} for all the computational details. 

Relative energies of the multiplet structure of the Ir$ _{2} $O$ _{9} $ dimer unit  
obtained from CASSCF + NEVPT2 (N-electron valence perturbation theory)~\cite{NEVPT2_CPL_2001} calculations are shown in Table~\ref{table:qc_wfc}. 
%At the CASSCF level, an active space of nine electrons in six orbitals  (three  $t_{2g}$ orbitals on each iridium) was considered. This multi-reference wave function calculation captures the static correlations that are very important in \BAIO. 
An active space of nine electrons in six orbitals  (three  $t_{2g}$ orbitals on each iridium) was considered in the CASSCF calculation which sufficiently captures the important static correlations (i.e. near degeneracies) in \BAIO. 
In the NEVPT2 calculation, the correlations involving all the neighboring occupied oxygen $2p$ and iridium $ 5s $, $ 5p $ orbitals as well as all the unoccupied orbitals are accounted for, accurately describing the O-2p to Ir-d charge transfer effects and other dynamic correlation effects.  
It is important to note that the intra- (Hund's coupling $J_H$) and inter-site ($U$) Coulomb interactions and the hybridization between different orbitals are included in the calculation, accurate within the basis-set limit. 
%
%The excitation energies were computed in a two-step procedure. First, a non-relativistic calculation for the lowest nine quartet ($ s = \frac{3}{2} $) and 24 doublet ($ s = \frac{1}{2} $) states (first column in Table~\ref{table:qc_wfc}, see caption) was performed. In a subsequent calculation, the spin-orbit matrix elements were computed in the basis of the 33 non-relativistic states and the resulting Hamiltonian is diagonalized to obtain the 84 SOC states as shown in the second column of Table~\ref{table:qc_wfc}.  
%

The lowest nine quartet ($ s = \frac{3}{2} $) and 24 doublet ($ s = \frac{1}{2} $) scalar relativistic states (first column in Table~\ref{table:qc_wfc}) are first computed and then are allowed to admix via the SOC, resulting in 84 states, see the second column of Table~\ref{table:qc_wfc}. It can be seen that the excitation energies obtained from CASSCF+NEVPT2+SOC calculations are in excellent agreement with the peaks observed in RIXS experiments, except for peak $ D $.
 %, see columns two and three. 
 This peak is related to the electron-hole exciton which is also observed in other iridate materials such as Sr$ _{2} $IrO$ _{4} $~\cite{Jungho_NatCom_2014,XingyeLu_OKedge_214_227_PRB_2018} and Na$ _{2} $IrO$ _{3} $~\cite{Ir213_rixs_gretarsson_2013}. 
 Such excitations are not considered in the current QC calculations\footnote{To simulate electron-hole exciton peaks one would need to calculate N-1 (removal of an electron) state calculations.}. 
 Further, our calculations reveal excitations from the $t_{2g}$ to $e_g$ manifold starting at 3.4 eV which correspond to RIXS peaks P and Q.
 
To elucidate the origin of these excitations, we first analyze the scalar-relativistic multiplet structure.
 When the two iridium ions in the dimer unit are in cubic environment ($ O_{h} $ symmetry), the low energy multiplet structure is a result of the interaction of the ground state $ ^2T_{1g}$ multiplet of the Ir$ ^{4+} $ ion~\cite{rhir214_vmk_14} and the $ ^{3}T_{1g} $ ground state term of the Ir$ ^{5+} $ ion. 
 In addition,  the lowest $^1T_{1g}$ and $^1E_{1g}$~\cite{Kush_d4_PRB_2018} singlet states contribute significantly to the low energy spin-orbit excitations~\cite{Kush_d4_PRB_2018}. 
 The resulting spectrum contains $ ^4T_{1g}, ^4T_{2g}, ^4A_{1g}, ^4E_{1g},  $  quartets and 11 doublet terms -- $ ^2T_{1g, 2g, 3g, 4g, 5g}$, $^2A_{1g, 2g, 3g}$, $^2E_{1g, 2g, 3g} $~\footnote{See direct product tables for $O_h$ point group representations, e.g. in Ref.~\cite{atkins_point_group_tables}}. 
 However, in \BAIO\ the Ir ions are enclosed in distorted octahedra resulting in low symmetry CFs and splitting of  the $ t_{2g} $ levels at each Ir ion~\footnote{For example, for pure trigonal distortions, the $ t_{2g} $ orbitals are split into $ a_{1g} $ and $ e_{g} $.}. 
 Further, the small Ir-Ir intra-dimer distance of 2.73 \AA\ in \BAIO\  (2.698 \AA\ in elemental iridium) may result in direct overlap of the Ir $d$ orbitals and the formation of bonding and antibonding states~\cite{Streltsov_PANS_2016}.
 Consequentially, the multiplet degeneracies in the spectrum are split. 
 
 To understand the formation of bonding and antibonding dimer orbitals in \BAIO, we plot the evolution of orbital energies as a function of Ir$_1$-Ir$_2$ intra-dimer distance ($d$) in Fig.~\ref{fig_energy_levels}(c). 
 The six levels for each $d$ correspond to the  CASSCF canonical orbital~\footnote{The canonical orbitals are eigenfunctions of the so-called (effective single particle) Fock operator.} energies of the six $ t_{2g} $-like orbitals in the Ir$_{2}$O$ _{9}$ dimer unit. % obtained directly from CASSCF calculations. 
 The colour variations of the energy levels represent the orbital compositions~\footnote{The orbital compositions are obtained from the L{\"o}wdin natural orbital decomposition.} from Ir$_1$, Ir$_2$ and O ions.
 Interestingly, for $d \ge 2.73$ \AA, we find 20\% and 13\% hybridization for Ir$ _{1} $ $5d$ -- O $2p$ and Ir$ _{2} $ $5d$ -- O $2p$, respectively, while there is negligible direct Ir$_1$-Ir$_2$ $d$-orbital hybridization.
The $ a_{1g} $ orbital of Ir$ _{1} $ contains 4.5\% contribution from $ a_{1g} $ orbital of Ir$ _{2} $ and vice versa. 
 For $d = 2.65$ \AA, a significant direct Ir$_1$-Ir$_2$ $ d $-orbital hybridization is observed. 
 We find this hybridization increasing up to 25\% for $d$ =$2.45$ \AA, resulting in large bonding -- antibonding energy separation, as seen in the corresponding orbital plots in Fig.~\ref{fig_energy_levels}(d) .
 Note that for $d=2.73$ \AA\, orbitals with predominantly Ir$_1$ character are at higher energies than those of Ir$_2$ character, reflecting different on-site orbital energies. 
 %For $d \ge 2.65$, 
 This is a direct consequence of the difference in the valence configurations of Ir$_1$ and Ir$_2$ ions and the Ir$_{1,2}$ $5d$ -- O $2p$ hybridization. 
  
 The effect of the low crystal filed symmetry at two different Ir ions can be estimated by computing the $t_{2g}$ splittings, $\delta$, at each of the Ir ions from restricted active space (RAS)~\cite{ras_roos_1991} calculations where the $d$ orbital occupation at the other Ir ion is constrained. 
 We find considerably large $t_{2g}$ splittings of $\delta_1=0.58$ and $\delta_2=0.60$ eV for Ir$_1$ and Ir$_2$, respectively. 
 Such large values compete with the SOC strength of $\sim 0.5$ eV of the Ir ions to considerably reduce the effect of SOC, thus resulting in the modification of the local spin-orbit multiplet structure. 
 
  The scalar-relativistic ground state realized in \BAIO\ is the double exchange $^4A_{1g}$ multiplet~\cite{Streltsov_prb_2017}, an orbitally non-degenerate high-spin quartet, with wave function   
 $^4\psi_0 = \alpha|d_1^4,d_2^5\rangle + \beta|d_1^3,d_2^6\rangle+\gamma |d_1^5,d_2^4\rangle $ with $\alpha^2=0.89, \beta^2=0.09$ and $ \gamma^2=0.02 $, where $d_i^n$ corresponds to $n$ electrons in Ir$_i$ $d$ orbitals. 
 The lowest $^2T_1$ doublet state is 40 meV higher with wave function weights $\alpha^2=0.85, \beta^2=0.11$ and $ \gamma^2=0.02 $.  
 We further find that the weight of $ |d_1^4,d_2^5\rangle $ configuration in all the excited multiplet wave functions is greater than 95\%. %, see SM~\cite{supmat}. 
 It is interesting to note that excluding all the configurations involving hopping of electrons from Ir$_1$ to Ir$_2$ and vice versa in the wave function preserve the spin-orbit spectrum except for an overall shift $\le$ 50 meV.  
 The double exchange ground state as well as the dominant contribution of $ |d_1^4,d_2^5\rangle $ configuration imply charge separation within the dimer units.
 Further, the natural orbital occupations obtained from the CASSCF calculations are close to 4 and 5 for Ir$_1$ and Ir$_2$ ions, respectively.  
 Thus, we conclude that the two Ir ions in \BAIO\  host different ionic states -- Ir$_1^{5+}$ and Ir$_2^{4+}$ -- which results in charge ordering within the dimers and the low energy excitations are strictly local to individual Ir ions and not among dimer orbitals.

 The SOC results in the admixture of all the 15 non-relativistic multiplet terms shown in Table \ref{table:qc_wfc}. 
 Addition of angular momenta of two $l_{\rm eff}$=1 ($l_{\rm eff, 1}$ = $l_{\rm eff, 2}$ = 1) sites with spins $s_1=1/2$ and $s_2 = 1$ gives rise to 84 effective total angular momentum ($J_{\rm eff}$) states. %, %with four doublet ($D$), five quartet ($Q$), three sextet ($S$) and one octet ($O$) multiplets, 
 %see SM~\cite{supmat} for details. 
 In \BAIO, due to the non-cubic CFs, all degeneracies are removed except for the Kramers doublet degeneracy. 
 From the analysis of the wave functions, we assign the peaks A-C to excitations from the Ir$_1$ $J_{\rm eff} =0$ to  $J_{\rm eff} =1$ states.
 The peak F and satellite feature H consists of excitations from Ir$_1$ $J_{\rm eff} =0$ to $J_{\rm eff} =2$ and  Ir$_2$ $J_{\rm eff} =1/2$ to  $J_{\rm eff} =3/2$ states that are split due to non-cubic CFs. 
 The peak G originates from excitations involving Ir$_1$ $J_{\rm eff} =0$ and $J_{\rm eff} =2$ states as well. 
 The peaks I-K are the result of simultaneous on-site excitations at Ir$_1$ and Ir$_2$ ions, see Fig.~S5 in SM~\cite{supmat}. 
 
 At the first sight, an insignificant Ir-Ir intra-dimer $d$-orbital hybridization in \BAIO\, might be surprising, even though the distance between Ir sites is close to that of Ir metal.
 However, %This is predominantly due to the difference in  
 due to the crystallographic in-equivalence of the two Ir ions in the dimer unit and the different O$_6$ arrangement, 
 the symmetry of split $t_{2g}$ orbitals at each of the Ir ions is very different and subsequently a little direct overlap is realized. 
 In fact, for dimer systems with structurally equivalent ions such as Ba$ _{3} $InIr$ _{2} $O$ _{9} $~\cite{Dey_IrDimer_PRB_2017}, we find a considerable hybridization resulting in delocalized dimer orbitals~\cite{vmk_BIIO}.  
 It would be interesting to characterize the local electronic structure in other face-sharing Ir-dimer compounds such as  Ba$ _{3} $ZnIr$ _{2} $O$ _{9} $ and Ba$ _{3} $ZrIr$ _{2} $O$ _{9} $~\cite{SAKAMOTO20062595} where the Ir dimer unit occupancy is eight and ten respectively.
  
  In conclusion, we have measured both Ir $L_3$ and O $K$-edge RIXS spectra and observed multiple spin-orbital excitations. Our {\it ab initio} quantum chemistry calculations reproduce very well the excitation spectrum up to 3.5 eV observed in the RIXS measurements. 
  We find charge ordering within the Ir-dimers with Ir$_1^{5+} {\rm (} d^{4} $) and Ir$_2^{4+} {\rm (} d^{5} $) configurations.  
  We have established a direct connection between the excitations in Ir$_2$O$_9$ dimer unit and those at individual Ir ions. 
  The appearance of multiple peaks is a direct consequence of strong non-cubic CFs originating from the distorted octahedral environment around the Ir ions. 
  In spite of small intra-dimer Ir-Ir distance, the direct $d$--$d$ hybridization is relatively weak and the bonding--antibonding splitting is negligible compared to the non-cubic CF splittings.  
  Alternatively, we find increased intra-dimer configuration mixing due to strong electron-electron interactions, particularly the $|d_1^3,d_2^6\rangle$ configuration stabilizing  the ground state. 
  This strongly supports nearly complete charge ordering within the Ir dimers in \BAIO\, and refutes the suggested formation of dimer orbitals~\cite{BIAO_wang_prl_2019}.
 Our results highlight the importance of minute details of the crystal structure to understand the electronic and magnetic properties of clustered Iridates and TM magnets in general and calls for re-investigating several already studied materials with accurate {\it ab initio} many-body calculations.
  Finally, we emphasize that the combination of RIXS and quantum chemistry calculations is an excellent tool to unambiguously decipher complicated electronic structures.  

\begin{acknowledgments}
V. M. K. and O. V. Y. would like to acknowledge funding from Swiss NSF NCCR MARVEL and the Sinergia grant NanoSkyrmionics CR- SII5 171003. The work at PSI is supported by the Swiss NSF through the NCCR MARVEL and the Sinergia network Mott Physics Beyond the Heisenberg Model (MPBH). X. L. acknowledges financial support from the European Community's Seventh Framework Programme (FP7/20072013) under Grant agreement No. 290605 (Cofund; PSI-Fellow). The work at BNU is supported by the National Natural Science Foundation of China under Grant No. 11734002 and 11922402. M. D. was partially funded by the Swiss National Science Foundation within the D-A-CH programme (SNSF Research Grant 200021L 141325). This research used resources of the Advanced Photon Source, a U.S. Department of Energy (DOE) Office of Science User Facility operated for the DOE Office of Science by Argonne National Laboratory under Contract No. DE-AC02-06CH11357.
\end{acknowledgments}
%\bibliography{Ref}

%merlin.mbs apsrev4-1.bst 2010-07-25 4.21a (PWD, AO, DPC) hacked
%Control: key (0)
%Control: author (0) dotless jnrlst
%Control: editor formatted (1) identically to author
%Control: production of article title (0) allowed
%Control: page (1) range
%Control: year (0) verbatim
%Control: production of eprint (0) enabled
%
\end{document}